\documentclass[12pt]{iopart}

\expandafter\let\csname equation*\endcsname\relax
\expandafter\let\csname endequation*\endcsname\relax

\usepackage{iopams}
\usepackage{amsmath}  
\usepackage{color}
\usepackage{lscape}
\usepackage{pdflscape}
\usepackage{ctable}
\usepackage[all]{xy}
\usepackage{soul}

\newcommand{\bes}{\begin{subequations}\bea}
\newcommand{\ees}{\eea\end{subequations}}
\newcommand{\be}{\begin{equation}}
\newcommand{\ee}{\end{equation}}
\newcommand{\bea}{\begin{eqnarray}}
\newcommand{\ba}{\begin{array}}
\newcommand{\eea}{\end{eqnarray}}
\newcommand{\ea}{\end{array}}

\newcommand{\detr}[1]{\det #1}

\newcommand{\la}{\big\langle}
\newcommand{\ra}{\big\rangle}

\newcommand{\Prob}{\mathrm{Prob}}

\begin{document}

\title{Generally covariant state-dependent diffusion}

\author{Matteo Polettini}

\address{Complex Systems and Statistical Mechanics, University of Luxembourg, Campus
Limpertsberg, 162a avenue de la Fa\"iencerie, L-1511 Luxembourg (G. D. Luxembourg)}
\ead{matteo.polettini@uni.lu}

\begin{abstract}
Statistical invariance of Wiener increments under $\mathrm{SO}(n)$ rotations provides a notion of \textit{gauge transformation} of state-dependent Brownian motion. We show that the stochastic dynamics of non gauge-invariant systems is not unambiguously defined. They typically do not relax to equilibrium steady states even in the absence of extenal forces. Assuming both coordinate covariance and gauge invariance, we derive a second-order Langevin equation with state-dependent diffusion matrix and vanishing environmental forces. It differs from previous proposals but nevertheless entails the Einstein relation, a Maxwellian conditional steady state for the velocities, and the equipartition theorem. The over-damping limit leads to a stochastic differential equation in state space that cannot be interpreted as a pure differential (It\=o, Stratonovich or else). At odds with the latter interpretations, the corresponding Fokker-Planck equation admits an equilibrium steady state; a detailed comparison with other theories of state-dependent diffusion is carried out. We propose this as a theory of diffusion in a heat bath with varying temperature. Besides equilibrium, a crucial experimental signature is the non-uniform steady spatial distribution.
\end{abstract} 


\section{Introduction}

Nonequilibrium thermodynamics is often modeled by means of stochastic differential equations (SDE's), describing possible trajectories in an open system's state space, and by diffusion equations for the evolution of ensembles \cite{sekimoto,seifert,esposito}. While the paradigm started by enhancing dynamical equations with an external source of noise, independent of the system's state, soon there arose the necessity of considering situations where the intensity and direction of the environmental disturbance depends on the state of the system \cite{hanggi,lau,klimontovich1}. From a different perspective, mathematicians had extensively studied SDE's on manifolds, where a state-dependent diffusion coefficient is often associated to a Riemannian metric (see Refs. \cite{hsu,stroock} for recent monographs). At the merge between the two approaches Van Kampen argued that ``The question of the existence and correct form of equations describing Brownian motion on a manifold cannot be answered by mathematics alone, but requires a study of the underlying physics'' \cite{vankampen}.

 \subsection{Scope and motivation} 

While statistical properties of Brownian motion are expected to only depend on the diffusion matrix and the physical drift, loosely speaking the ``square root'' of the diffusion matrix appears in a SDE. In one dimension this is unambiguous, while in $n$ dimensions there are many different ways to take such square root, differing one from another by local internal rotations. We will require the physical description to be independent of this choice. When this doesn't occur, for a given diffusion matrix one can write down SDE's that have very different dynamics and thermodynamics. In particular most choices lead to nonequilibrium steady states even in the absence of external forces: An unobservable statistical property oddly becomes a nonequilibrium drive. 

Taking up Van Kampen's invitation, in this paper we make a proposal for state-dependent diffusion equations based on the physical symmetry principle of General Covariance, by which we mean:
(1) Covariance under external coordinate transformations; (2) Invariance under local internal rotations (gauge invariance). The covariance of nonequilibrium statistical mechanics under coordinate transformations has been extensively studied by Graham \cite{graham} (also Ref.\cite{muratore} for applications to semiclassical methods in field theory). Departing from Graham's analysis, we further assume that statistical properties of the theory are invariant under a class of internal gauge transformations under rotations of the noise components, taking inspiration from the frame field approach to General Relativity. However, we do \textit{not} derive relativistic equations for diffusion in strong gravitational fields.

Recently we witness an increasing interest in geometric methods in nonequilibrium statistical mechanics  \cite{geometric1,crooks,ren,geometric2,qian,polettini}, with emphasis often given to the equilibrium vs. nonequilibrium character of steady states. This paradigm will allow us to better appreciate the physical nature of our construction. It is our conviction that the theory is relevant to diffusion in environments with nonuniform temperature, as we briefly anticipate in the next paragraph and  argue more thoroughly in a parallel publication \cite{temp}.

\subsection{Models and results}

In this paper we consider stochastic models of systems subject state-dependent diffusion, viscous damping and no external forces, unless otherwise stated. We do not consider velocity-depedent diffusion, as is done in Ref.\cite{klimontovich1}. To this purpose second order Langevin-type equations serve the purpose better than do first order SDE's. In fact, the latter are plagued with the problem of interpretation, which makes ``vanishing forces'' a fuzzy concept. Moreover, second order equations go onstage on a geometrically neat and tidy tangent bundle.

In one dimension, the following equation with state-dependent diffusion coefficient and vanishing environmental forces is frequently encountered \cite{lau,sanchodurr,sancho}
\be
\ddot{x}_t = - \gamma(x_t)  \dot{x}_t + \sqrt{2\sigma}\, e(x_t) \zeta_t, \label{eq:lan1D}
\ee
where $\sqrt{2} \, \zeta_t$ is Gaussian white noise, $\sigma(x) =  \sigma e(x)^2$ is a state-dependent diffusion coefficient and $\gamma(x)$ is state-dependent viscosity. We will later refer to this theory as ``state-dependent viscosity''. Assuming the state-dependent Einstein relation $\gamma(x) = \beta \sigma(x)$, the theory leads to a Maxwellian steady state $P^\ast(x,v) \propto \exp - \beta v^2/2$, with uniform spatial density. Experimental evidence by  Lan\c con \cite{lancon}, Volpe \cite{volpe,fog} and coworkers seems to indicate that Eq.(\ref{eq:lan1D}) describes the motion of an overdamped colloidal particle when the bath has uniform temperature, while displaying a nonuniform density due to hydrodynamic interactions with confining walls.

Less studied \cite[ \S 1.3.2.1, \S 4.1.2.2]{sekimotobook} \cite{matsuo,bringuier,celani} is the following equation 
\be
\ddot{x}_t = - \gamma \dot{x}_t + \sqrt{2\sigma}\, e(x_t) \zeta_t. \label{eq:temp}
\ee
It was originally proposed by Van Kampen \cite{vankampenIBM} to realize a theory settling Landauer's physical intuitions \cite{landauer} about systems with varying temperature, and therefore we will later refer to this theory as ``state-dependent temperature''. In fact, it is obtained from the state-independent Langevin equation by simply replacing $\beta$ with $\beta(x) = e(x)^2\gamma/\sigma$. It differs from Eq.(\ref{eq:lan1D}) in one respect: The viscosity $\gamma$ is not itself state-dependent. It leads to a nonuniform steady distribution of particles $p^\ast(x) \propto \beta(x)$, in accordance with the phenomenon of thermophoresis which drives particles along the temperature gradient, from hotter to colder zones.

We advance yet another equation: 
\be
\ddot{x}_t = - \gamma \dot{x}_t +  \left[ \partial_x \log e(x_t) \right] \dot{x}_t^2 + \sqrt{2\sigma}\, e(x_t) \zeta_t. \label{eq:genlan1D}
\ee
As the latter theory,  it involves a state-independent viscosity. However, an additional quadratic term in the velocity appears --- somewhat analogous to Rayleigh's drag force. The theory accomodates Einstein's relation and a Maxwellian steady conditional probability $\pi^\ast(v|x) = P^\ast(x,v)/ p^\ast(x)$ for velocities at a given position, while the spatial density  is generally  not uniform,  dispensing with the equal-a-priori postulate. In Ref.\cite{temp} we argue that this might be a valid alternative theory for systems with nonuniform temperature, even more adherent to Landauer's thermodynamic picture. In the present paper we discuss the geometrical properties of the theory and of its $n$-dimensional generalization.

In the overdamping limit $\gamma,\sigma \to \infty$ at fixed $\beta$, the viscous damping term is responsible for the reduction of Langevin-type equations to first order equations over state-space, where the inertia $\ddot{x}$ of the particle can be neglected. Notoriously, the latter are subject to antique \cite{vankampen2,klimontovich2}
and ongoing \cite{lau,sancho,lancon,volpe,fog,sokolov,yuan,mannella} debates on the ``correct'' interpretation of the stochastic differential: It\=o,  Fisk-Stratonovich, H\"anggi-Klimontovich, or else? Since different stochastic differentials are mapped one into another by adding effective drift terms, the problem of interpretation is equivalent to the choice of drift in a given fixed convention; throughout this paper we choose to work with It\=o calculus. We talk of \textit{pure} SDE when the equation has no drift in some convention. Remarkably, the generally covariant first-order SDE stemming from our theory is not pure, as it requires and effective drift in all conventions, but for the one-dimensional case, where it reduces to the Stratonovich differential. It rather coincides with the  mathematicians' \textit{definition} of Brownian motion on a manifold.

One remarkable property of our construction is that the steady state of the generally covariant reduced Fokker-Planck equation always satisfies detailed balance, with vanishing steady currents, while other approaches might lead to nonequilibrium steady states even in the absence of environmental forces, with steady currents driven by  inhomogeneity of the diffusion coefficient. In particular this is always the case for all pure stochastic differentials.

\subsection{Plan and notation}
 
The paper is organized as follows. In Sec. \ref{gencovlan} we discuss the statistical meaning of gauge invariance under rotations and derive the generally covariant Langevin equation, the corresponding Kramers generator and its steady state. Sec. \ref{firstorder} is devoted to the Fokker-Planck equation and to the corresponding first order SDE deriving from the overdamping limit, which is carried out in \ref{appendix2} (for a slightly more general theory). We also define covariant thermodynamic forces, currents, entropy and entropy production. In Sec. \ref{comparison} we compare our results with other theories of state-dependent diffusion and provide comparative tables in \ref{compa}. Finally we draw conclusions in Sec. \ref{conclusions}.

In the following, $x,x'$ are local coordinates on an $n$-dimensional manifold $M$, which is assumed to be compact without boundary to avoid problems with conservation of probability; $x_t,x'_t$ are coordinates of a stochastic trajectory; Explicit dependences on coordinates and time are often omitted; Indices $a,b,c$ and $i,j,k$ label respectively coordinate and euclidean directions; The Einstein convention on index contraction is used; Spatial derivatives are abbreviated with $\partial_a = \partial / \partial x^a$; Tensor indices are raised and lowered with the metric, $X_a = g_{ab} X^b$, internal indices with the euclidean metric $X_i = \delta_{ij} X^j$. Fields and probabilities are assumed to be smooth, so that there is no nonequilibrium driving due to topological defects such as sinks and sources.

\section{\label{gencovlan}Brownian motion}

\subsection{\label{gauge}Gauge invariance under rotations}

We work with $n$ independent sources  $\zeta^i_t$ of Brownian white noise. They are the formal time derivatives of $n$ independent Wiener processes $(W^i_t)_t$. Wiener increments $dW^i_t = \zeta^i_t dt$ have null average, are uncorrelated at different times and in different internal directions,
\be
\la dW^i_t \ra  = 0 , \qquad 
\la dW^i_t  \, dW^j_{t'}\ra = \delta^{ij} \delta(t-t') dt, \label{eq:statprop}
\ee
and are normally distributed  with covariance $\delta^{ij} dt$:
\be
\Prob \, ( dW^i_t ) \propto \exp {- \frac{\delta_{ij} dW^i_t dW^j_t}{2dt}}.  \label{eq:statdist}
\ee
All three properties are invariant under rotations of the Wiener increments, $dW^i \to R^i_{\phantom{i}j} dW^j$, where $R_{ij}$ a special orthogonal matrix,
\be
R^i_{\phantom{i}j} R^k_{\phantom{k}l} \delta_{ik} = \delta_{jl},
\ee
with unit determinant $R = \detr{(R_{ij})_{i,j}} = 1$. The reader should pay attention to the collocation of indices to avoid confusion between a matrix and its transpose/inverse. For example, $R^{\phantom{i}i}_{j} R^j_{\phantom{j}k} = \delta^i_k$ while $R^{i}_{\phantom{i}j} R^j_{\phantom{j}k} = (R^2)^i_k$. Invariance  under global rotations is obvious for Eq.(\ref{eq:statdist}) and is easily proven for Eqs.(\ref{eq:statprop}) from the linearity of the average.

The gauge principle suggests to consider the analogous local symmetry. The intuitive picture is as follows. Wiener increments are attached to each point of state space. As a Brownian particle passes by  $x_t$ at time $t$  it receives a kick. While single trajectories do change if we rotate the Wiener increment \textit{at that point} by a rotation matrix $R^i_{\phantom{i}j}(x_t)$, the statistics of kicks remains unchanged. In fact, linear combinations of independent multivariate normal variables have multivariate normal distribution, with average
\be \la R^i_{\phantom{i}j}(x_t) dW^j_t    \ra = \la R^i_{\phantom{i}j}(x_t)\ra \la dW^j_t \ra = 0, \label{eq:factor1} \ee
and covariance matrix
\be
\la R^i_{\phantom{i}k}(x_t) dW^k_t   \,R^j_{\phantom{j}l}(x_t) dW^l_t \ra = \la R^i_{\phantom{i}k}(x_t) R^j_{\phantom{j}l}(x_t) \ra \la dW^k_t   dW^l_t \ra
= \delta^{ij} \delta(t-t') dt . \label{eq:factor2}
\ee
In both cases we used the properties of It\=o calculus, which we conventionally work with. More precisely, any product $h(x_t) dW_t^i$ of a state function by a stochastic increment requires an interpretation about when the state function is evaluated within the time interval. The interpretation stems from the underlying discretization procedure. The trajectory is sampled at discrete times $(t_{\kappa})_\kappa$, when it occupies states $(x_{\kappa})_\kappa$. At intermediate times $t \in [t_{\kappa},t_{\kappa+1})$ one approximates $h$  according to $h(x_t) = h(x_{\kappa} + \alpha (x_{\kappa+1} - x_{\kappa}))$, where: $\alpha = 0$ (It\=o) has a causalistic, non-anticipatory character;  $\alpha = 1/2$ (Fisk-Stratonovich) is middle-point and preserves the rules of calculus; $\alpha = 1$ (H\"anggi-Klimontovich) has a finalistic character. We emphasize that the choice of It\=o calculus is purely conventional and does not affect the generality of the results. More importantly, this choice is \textit{not} related to the dispute that we hinted at in the introduction. 

The advantage of It\=o calculus is that the statistical properties of Wiener increments are independent of the state $x_t$ where they are nested, hence we are allowed to factor as we did in Eqs.(\ref{eq:factor1},\ref{eq:factor2}). What these equations then tell us is that the rotated Wiener increments have the same statistical properties as the starting ones: They are statistically indistinguishable. This is a gauge transformation in that it acts on the internal properties of the system --- we do not measure the stochastic increments themselves --- and it depends differentiably on the state.

Since Wiener increments enjoy this symmetry, we expect that any theory whose statistical properties are dictated by the statistics of Wiener increments also enjoys this symmetry. As we will point out, just like theories that are not coordinate covariant depend on the particular coordinates chosen, theories which do not satisfy gauge symmetry have very different properties according to how the symmetry is made to break. Gauge invariance of overdamped Brownian motion has also been recently discussed in Ref.\cite{muratore2}.

\subsection{Generally covariant Langevin equation}

While the  the manifold of states $M$ needs not be physical space, for sake of visualization the Langevin equation
\be
\dot{x}^a = v^a, \qquad \dot{v}^a  = -   \gamma v^a + F^a(x,v) + \sqrt{2\sigma}\, e^a_i(x) \zeta^i , \label{eq:langevin}
\ee
is thought to describe the motion of a Brownian particle with mass $m$. The trajectory is driven by an effective drift $F^a$, is damped by a viscous force  $- \gamma \dot{x}^a$, and receives stochastic kicks from the $\zeta^i_t$'s. The \textit{frame} $e^a_i$ mixes $n$ independent Wiener increments to yield $n$ coordinate increments. It determines intensity and direction of random kicks, in a way that  depends on the state visited by the trajectory. Correlations are characterized by the diffusion matrix
\be
g^{ab}(x) = e_i^a(x) e_j^b(x)  \delta^{ij} . \label{eq:diffmat}
\ee
We assume that $(e^a_i)_{i,a}$ is an invertible matrix, that is, the $n$ frame vectors $e_i = (e^a_i)_{a}$ are linearly independent. Its inverse $(e_a^i)_{i,a}$ is called the coframe. It follows that the diffusion matrix is positive definite, with inverse $(g_{ab})_{a,b}$ and inverse determinant $g = \det (g_{ab})_{a,b} > 0$.

We now perform an invertible coordinate transformation $x \to x'(x)$ and, by the principle of covariance, we assume that the trajectory $x'_t = x'(x_t)$ obeys an equation in the same form as Eq.(\ref{eq:langevin}):
\be
\dot{x}^{a'} = v^{a'}, \qquad \dot{v}^{a'}  = -   \gamma v^{a'} + F^{a'} + \sqrt{2\sigma}\, e^{a'}_i \zeta^i. \label{eq:translangevin}
\ee
In theory, we should resort to the rules of It\=o calculus to evaluate differentials; in practice, for this kind of equation stochastic and standard calculus coincide, and there is no issue concerning the interpretation of the stochastic differential. From the first transformed equation $\dot{x}^{a'} =  v^{a'}  \partial_a x^{a'}$ we read off the transformation law for the velocity  $v^{a'} = v^a \partial_a x^{a'}$, which  behaves like a vector sitting in the tangent space at $x$. We then take the stochastic differential of $v^{a'}$, using the (It\=o-)Leibniz rule
\be
dv^{a'} = v^a d\partial_a x^{a'} + dv^a\partial_a x^{a'} 
= \big( \partial_a\partial_b x^{a'} v^a v^b+ \partial_a x^{a'}   F^a \big)  dt  + \sqrt{2}\,  \partial_a x^{a'}  e^a_i  \zeta^i dt,
\ee
which leads to
\be
e_i^{a'} =  \frac{\partial x^{a'}}{\partial x^a} e_i^a, \qquad
F^{a'} =  \frac{\partial x^{a'}}{\partial x^a} \ F^a + v^a v^b \frac{\partial^2 x^{a'}}{\partial x^a \partial x^b}  . \label{eq:trans}
\ee

The frame's components, at fixed $i$, transform like vectors, while the inverse diffusion matrix is twice contravariant, $g_{ab} = g_{a'b'} \partial_a x^{a'} \partial_b x^{b'} $. Therefore, the latter bears the properties of a metric, with invariant volume element $\sqrt{g} \, dx$. Since $g_{ab}e^a_i e^b_j = \delta_{ij}$, the frame vectors $e_i$ form a basis of orthonormal vectors for the tangent space at $x$, whence the interpretation as orthonormal reference frame, point-by-point over the manifold. In general, such reference frames are anholonomic (noncoordinate), meaning that the matrix $e_i^a$ is not the Jacobian of a coordinate transformation that makes $e^a_i = \delta^a_i$ and $g_{ab}$ the flat euclidean metric, but at one point on the manifold at a time. The frame makes its appearance in General Relativity, where it is known as vielbein (counting  dimensions in german, viel = zwei, drei, vier, etc.). For example, the vierbein maps the local observer's 4D flat Minkowsky metric --- three sticks and a clock, following Einstein's physical intuition --- into a semi-Riemmanian metric on the spacetime manifold.

As is well known, the inhomogeneity developed by the drift term in Eq.(\ref{eq:trans}) qualifies it as 
\be
F^a_{1}  = - \Gamma^a_{bc} v^b v^c,\label{eq:geodesicterm}
\ee
where $\Gamma^a_{bc}$ are the Christoffel symbols of the Levi-Civita connection, \textit{i.e.} the unique torsionless connection compatible with metric $g_{ab}$,
\be
2\Gamma^a_{bc} = g^{ad} \left(\partial_b g_{dc} + \partial_c g_{db} - \partial_d g_{cb} \right).
\ee
However, this is not the only viable choice, as
\be F^a_{2} = v^c v^b e^i_b \partial_c e^a_i \ee
produces the same inhomogeneous term. As we will see later, in the overdamping limit the latter leads to Graham's theory while the former leads to the mathematicians' Brownian motion on a manifold. 

For a sensible choice between the two, we invoke the principle of gauge invariance. We assume that the theory is invariant under internal $\mathrm{SO}(n)$-transformations of the reference frames, acting on the $i,j,k$ indices
\be
e^a_i(x) \to  \tilde{e}^a_i(x) =  e^a_j(x) R_{\phantom{j}i}^j(x),  \qquad
e_a^i(x) \to  \tilde{e}_a^i(x) =   e_a^j(x) R^{\phantom{j}i}_j(x),   \label{eq:gaugetransframe} 
\ee
where the transormation law for the coframe follows  from $\tilde{e}^a_i \tilde{e}_b^j = \delta^a_b$.
Notice that the metric is invariant under local coordinate transformations, $\tilde{g}_{ab} = g_{ab}$, and so are the Christoffel coefficients. In General Relativity, gauge transformations have the physical meaning of rotations of the observer's orthogonal reference frame. As we argued in the dedicated paragraph, $\zeta_t^i$ and $\tilde{\zeta}^j = R_{\phantom{j}i}^j(x) \zeta_t^i$ are statistically indistinguishable, hence  the stochastic term is invariant under $e^a_i \zeta^i_t \to \tilde{e}^a_i \zeta^i_t = e^a_i \tilde{\zeta}^i_t$.

Given that the stochastic term is statistically preserved by gauge transformations, it remains to check that the deterministic term is. As a rule of thumb, it should be expressible only in terms of the metric and of its derivatives. This is the case for $F_1$, while it is not for $F_2$. In fact, after a gauge transformation:
\be
\delta F_1^a = 0, \qquad  \delta F_2^a = e_k^a e^j_b v^b v^c  R^{\phantom{j}i}_j  \partial_c R_{\phantom{k}i}^k .  
\ee
Therefore we choose $F_1$, obtaining
\be \dot{v}^a + \Gamma^a_{bc} v^b v^c = - \gamma v^a + \sqrt{2\sigma}\, e^a_i \zeta^i . \label{eq:gencovlan} \ee
In the deterministic case $\gamma = \sigma = 0$ one recognizes the geodesic equation. Eq.(\ref{eq:gencovlan}) is our proposal for a Langevin equation with state-dependent diffusion matrix. On a flat manifold, in coordinates where the metric is euclidean and the Christoffel coefficients vanish, it returns the ordinary Langevin equation.

At the time of submitting, the author came to knowledge that Eq.(\ref{eq:gencovlan}) can already be found in a work by Kleinert and Shabanov \cite{kleinert}, who discussed its generalization to connections with torsion. Very recently it has also drawn the attention of Castro-Villareal and coworkers \cite{castro2}, who derived it using an extrinsic, rather than intrinsic, approach.

\subsection{\label{kramers}Kramers equation and its steady state}

While the $n$-dimensional process $(x_t)_t$ is not a Markovian diffusive process, the $2n$-dimensional process $z_t = (x_t,v_t)$ on the tangent bundle  $TM$ is. By a standard procedure, the Kramers equation is obtained by writing down the diffusion equation for the probability density $P_t(z)$ and expressing it in in terms of $(x,v)$: 
\be
\dot{P} = - v^a \frac{ \partial  P}{\partial x^a}  + \frac{ \partial }{\partial v^a} \big[ \big( \Gamma^a_{bc}v^b v^c + \gamma v^a \big) P \big] + \sigma g^{ab}   \frac{ \partial^2 P}{\partial v^a \partial v^b} . \label{eq:covlanfp}
\ee
The steady state is given by
\be
P^\ast(x,v) =\frac{g(x)}{N} \exp {- \frac{\gamma}{2\sigma} g_{ab}(x) v^a v^b} .\label{eq:steadylangevin} 
\ee
It can be checked by direct substitution; the calculation entails an interesting balancing of Christoffel terms $\Gamma_{abc} v^a v^b v^c$ and $2 \Gamma^b_{ab}  v^a$ from $v$ to $x$ derivatives.

Since the metric has physical dimensions $[x]^{-2}$, the viscosity and the diffusion constant have dimensions respectively $[\gamma] =[t]^{-1}$ and $[\sigma]=[t]^{-3}$.  We define an energy,
\be
H(x,v) =  \frac{m \lambda_c^2}{2} \, g_{ab}(x) v^a v^b, \label{eq:energy}
\ee
where we introduced the Compton wavelength of the Brownian particle  $\lambda_c = 2\pi \hbar/mc$ for sake of dimensional consistency. In \ref{appendix1}  it is shown that the average energy $E_t = \langle H(x_t,v_t) \rangle$ is a Lyapunov functional, exponentially decreasing to the steady value $n \beta^{-1}/2$, where
\be
\gamma/ \sigma =  \beta (2 \pi \hbar)^2 /  mc^2. \label{eq:einstein}
\ee
The identification of $\beta$ with an inverse invariant temperature is suggested by the equipartition theorem at the Gibbs steady state $P^\ast \propto \exp - \beta H$, yielding the Einstein relation. Upon coordinate transformations, the energy is a scalar, the temperature is invariant and the probability density transforms like a squared volume density (like $g$), consistently with the fact that $P_t(x,v) dx \wedge dv$ ought to be an invariant volume element to preserve normalization.

Finally, we calculate the steady spatial density by integrating over velocities:
\be
p^\ast(x) = \int_{-\infty}^{+\infty} P^\ast(x,v) \, dv
 = \frac{\sqrt{g(x)}}{\int_{M} \sqrt{g(x')} dx'}  \label{eq:ss}
 \ee
 where $N = ( mc^2/ 2 \pi \beta \hbar^2)^{n/2} \int_{M} \sqrt{g(x')} dx' $. Brownian particles tend to occupy space according to the volume element. More probable states occur where noise is less intense. In fact, if we modulate the intensity of noise without modifying the angles between frame vectors, by performing a rescaling  $e_i  \to \lambda_i e_i$ (no index contraction), the determinant of the metric scales like $g \to g/\prod_i \lambda_i$. This is physically consistent, as more intense noise kicks    particles further away, making permanence at a state briefer. Modifying the angles, $g$ increases when the frame vectors are made less and less linearly independent, shooting to infinity when $(e_i^a)_{i,a}$ becomes degenerate.

 The conditional probability density $\pi^\ast(v|x) = P^\ast(x,v)/p^\ast(x)$, describing how  velocities in the tangent space $T_x M$ distribute at a fixed $x$, is Maxwellian. However, since  $x$ and $v$ terms in $P^\ast$ do not factor, the velocity probability density $\Pi^\ast(v) = \int_M P^\ast dx$ will generally not be Gaussian. This is a distinctive signature of the theory.

\section{\label{firstorder}Overdamped dynamics}

\subsection{Fokker-Planck equation}

The overdamping (Smoluchovski) limit of the Langevin equation is achieved by sending $\gamma,\sigma \to \infty$ at fixed $\beta$, in which case the damping effect of the viscous force is so strong that the inertial term $\dot{v}^a$ drops. The limit must be taken with great care; details of a rigorous derivation are reported in  \ref{appendix2}, where the Kramers equation is shown to reduce to the following Fokker-Planck equation
\be
\dot{p}_t  =  \partial_a \left[ g^{ab}   \left( \partial_b p_t -   p_t \, \partial_b \log \sqrt{g}\right)\right] \label{eq:covfp}
\ee
for the spatial probability density
\be
p_t(x) = \int dv\, P_t(x,v).
\ee
Introducing the covariant derivative $\nabla_{\!a}$ associated to the Levi-Civita connection, defined by $\nabla_{\!a} X^b = \partial_a X^b + \Gamma_{ac}^{b} X^c$, Eq.(\ref{eq:covfp}) can be recast in manifestly covariant form
\be
\dot{\rho} =\Delta_M \rho -  \nabla_{\!a} \left( f^a \rho \right), \label{eq:covfpe}
\ee
where $\rho_t = p_t /\sqrt{g}$ is a scalar  and $\Delta_M = \nabla_{\!a} \nabla^a$ is the Laplace-Beltrami operator. For sake of generality we momentarily accounted for an environmental force vector $f^a$. In deriving Eq.(\ref{eq:covfpe}), we used the fact that the covariant derivative of a scalar coincides with its partial derivative and that the covariant gradient is given by $\nabla_{\!a}  X^a = \sqrt{g}^{\;-1} \partial_a (\sqrt{g} \, X^a)$.  When  $f^a = 0$, Eq.(\ref{eq:covfpe})  returns the  well-known definition of Brownian motion on a manifold in the specialized mathematics literature (up to a factor $1/2$) \cite[Ch.3]{hsu} \cite[\S 4.2.2]{stroock}; It was also considered by Batrouni et al. in Ref. \cite{batrouni} with particular interest to the stochastic quantization of lattice gauge theories, and by Castro Villarreal \cite{villarreal}, who worked out the small-time/low-curvature dynamics. 

\subsection{Smoluchovski equation}

The SDE that corresponds to Eq.(\ref{eq:covfp}) is given by
\be
\dot{x}^a = f^a + f^a_{\mathrm{eff}}  + \sqrt{2}\,  e_i^a \zeta^i, \label{eq:gencov}
\ee
where again we included the environmental force vector, and the effective drift reads
\be
f^a_{\mathrm{eff}}   = -g^{bc} \Gamma^a_{bc} = \partial_b g^{ab} + g^{ab} \partial_b \log \sqrt{g}.  \label{eq:christoffel}
\ee
Eq.(\ref{eq:gencov}) can also be derived from first principles in much the same way as the generally covariant Langevin equation, keeping into account that by the rules of It\=o calculus the effective drift transforms like
\be  f^{a'}_{\mathrm{eff}} = \partial_a x^{a'}(  f^a_{\mathrm{eff}} + g^{ab} \partial_a \partial_b  x^{a'}). \ee
In short, one performs a Taylor expansion of $x'(x_t)$ to second second order in $dx_t$, plugs Eq.(\ref{eq:gencov}) in and retains terms of order $dt$, keeping in mind that squared Wiener increments $(dW_t^i)^2$ equal $dt$ with certainty.

Interestingly, when $f^a = 0$, for no value of $\alpha$ does Eq.(\ref{eq:gencov}) coincide with a pure SDE
\be
\dot{x}_t =  \sqrt{2} \, e^a_i(x_t + \alpha \, dx_t) \zeta^i_t, \label{eq:conventions}
\ee
as there always appears an effective drift term, but in one dimension, where it reduces to the Stratonovich differential.
We notice in passing that the It\=o differential leads to a gauge invariant, but not coordinate-covariant SDE, while the converse occurs for  the Stratonovich convention. In all cases, upon either a coordinate or a gauge transformation, such differentials usually gain an additional effective drift.  More interestingly, pure SDE's in Eq.(\ref{eq:conventions}) usually lead to nonequilibrium steady states, as we sparsely argue in Sec. \ref{comparison}.

\subsection{Geometric insights}

The promotion of global symmetries to local ones by virtue of the gauge principle usually entails the appearance of a new field, which compensates the transformation law and introduces a new interaction. The gauge field for $\mathrm{SO}(n)$ transformations is the \textit{spin connection} $\omega_{aij}$, which generates an infinitesimal rotation as one displaces vectors along a given direction, \textit{viz.} $(\omega_{aij})_{i,j}$ is an element in the Lie algebra $\mathfrak{so}(n)$ of skew-symmetric $n \times n$ matrices. Here we show how our theory writes nicely in terms of the spin connection and of the closely related anholonomicity vector. This is the only truly geometrical part of the paper; the reader might safely skip it. Nevertheless, we maintain the general inaccurate tone depicting geometry as index gymnastics.  

Since fields are tensors, we should hereby resort to Stratonovich calculus, which preserves the rules of differential calculus. Eq.(\ref{eq:gencov}), with $f^a = 0$, now reads
\be
\dot{x}^a = - \delta^{ij} e^b_i \partial_b e^a_j   - g^{bc} \Gamma^a_{bc} +  \sqrt{2}\,  e_i^a \circ\zeta^i =  - \omega^a + \sqrt{2}\,  e_i^a \circ\zeta^i, \label{eq:stratogamma}
\ee
where $h(x_t) \circ \zeta_t^i = h(x_t + dx_t/2)\zeta_t^i$ is the Stratonovich differential. In the first passage we added a drift term to switch convention. After some manipulation one obtains the second equation, where
\be  \omega_a =  e^b_i (\partial_a e_b^i - \partial_b e^i_a)  \label{eq:anholonomicity} \ee
is called the \textit{anholonomicity (co)vector}. After a gauge transformation the anholonomicity vector gains an inhomogeneous term which on average balances exactly the one developed by the  Stratonovich differential
\be
 \delta  \left\langle\omega_a \right\rangle =   - \left\langle e^k_a  e^c_j R_i^j  \partial_c R^i_k \right\rangle = \delta  \left\langle  \sqrt{2}Ê\, e^a_i(x_t) \circ \zeta_t^i  \right\rangle .  \label{eq:gaugetrans}
\ee
Notice that the anholonomicity vector has the typical transformation law $G^{-1} \partial_a \, G$ of gauge fields, where $G$ is a group element. Therefore, the anholonomicity vector is the gauge field of generally covariant stochastic thermodynamics. It will be present in all expressions where rotation invariance requires adjustment. In a way, the anholonomicity vector introduces an effective interaction between the Brownian particle and the environment which bends its motion.

Let's see how the theory writes in terms of the spin connection. The latter is used to displace objects with internal indices along the manifold rigidly, according to $\nabla_{\!a} X^i = \partial_a X^i + \omega^i_{aj}X^j$, while spatial indices are displaced with the Levi-Civita connection.  A condition for a metric-preseving connection is that the covariant derivative of the coframe shall vanish, 
\be
\nabla_a e_b^i = \partial_a e_b^i - \Gamma_{ab}^c e_c^i + \omega^i_{aj} e_b^j  = 0.
\ee
Antisymmetrizing, and recalling that the Levi-Civita connection is torsionless, $\Gamma^c_{[a,b]} = 0$, we can express the anholonomicity vector as
\be
\omega_a = \omega^i_{bj} e_a^j  e^b_i,
\ee
where we used $\omega_{ai}^i = \delta^{ij} \omega_{aij} =  0$.  Finally we can write Eq.(\ref{eq:stratogamma}) in a particularly simple form in terms of the spin connection and of the the velocity  $v^i_t = e_a^i(x_t) \circ \dot{x}_t^a$:
\be
v^i +   e^a_j \omega^{ji}_{a} = \sqrt{2}\, \zeta^i .
\ee

Let us briefly discuss the geometric significance of the anholonomicity vector. It measures to what extent the orthogonal basis fails to be a coordinate basis. If there exists a collection of potentials  $u^i$ such that $e_a^i= \partial_a u^i$, then $\omega^a$ vanishes. Performing coordinate transformation $x^a \to u^i(x)$, the transformed metric is easily seen to be euclidean. In this case, the manifold is  flat, and it has vanishing Riemann curvature. In general, the Riemann curvature of a manifold does not vanish, so that for such systems it is impossible to find coordinates where the anholonomicity vector vanishes. We skip the case where the manifold is only locally flat. We only observe that (local) flatness has no thermodynamic relevance to our theory, as far as we can appreciate.

\subsection{Covariant entropy and entropy production}

Entropy and entropy production are central to macroscopic thermostatics and thermodynamics. We give covariant analogues in first-order theoy, while a complete thermodynamic treatment is deferred to a future publication. Covariant equilibrium thermostatics has also been recently discussed in Ref.\cite{muratore2}.

From an information-theoretic point of view,  the uniform distribution maximizes the Gibbs-Shannn entropy, a measure of ignorance with respect to a reference prior, given by the uniform distribution itself. The Gibbs-Shannon entropy notoriously develops an inhomogenous term when one keeps the uniform prior fixed through a coordinate change, $\delta S = - \int p \log \det  \partial x' \, dx$. It can be avoided by transforming the prior as well, which will therefore not be uniform in the new coordinates.

This to say that it does not make much sense to pick the uniform prior in some preferred coordinates in the first instance. Our theory comes with its own preferred prior, namely the volume element $\sqrt{g}$. A  natural covariant candidate as a measure of \textit{excess information} is then
\be D_{\sqrt{g}} = \int  dx \, \sqrt{g} \, \rho \log \rho, \ee
where $\rho = p/\sqrt{g}$ is a scalar. The above expression is relative entropy with respect to $\sqrt{g}$. Though not uniform (a matter of concern for some authors \cite{tupperyang}), $p^\ast$ is analogous to the microcanonical ensemble, with equal volumes in state space having equal probabilities. There is an obvious circularity in this argument, since the choice of $\sqrt{g} \, dx$ as the preferred volume element is subjective, as is equiprobability. In fact, $\sqrt{g}$ bears more than a resemblance with Jeffrey's prior in Bayesian inference, whose objective vs. subjective character is still debated \cite{objsubj}. 

The author recently argued that subjective priors have full citizenship in NESM \cite{polettini}. A change of prior does modify the thermo\textit{statics} of the system but not its thermo\textit{dynamics}, which is fully encoded in the concept of entropy production, which we now introduce. Let us light up external forces for sake of generality. We introduce  covariant current and affinities,
\be
J^a = f^a \rho - \nabla^a \rho, \qquad
A_a = f_a -   \nabla_b \log \rho,
\ee
and the entropy production
\be
\sigma
= \int dx\, \sqrt{g} \, J^a A_a  = \int p(dx) \, g_{ab} A^a A^b = \left\langle A^2  \right\rangle.
\ee
The equalities on the r.h.s. show that the entropy production is non-negative. It vanishes at the steady state when the condition of detailed balance holds, that is when there exists a potential $\phi$ such that $f_a = - \nabla_a\phi$, with steady state $\rho = \sqrt{g} \exp - \phi$. Notice that our free theory, with $f_a = 0$, satisfies detailed balance. We mention that a covariant necessary condition for detailed balance is
\be
\nabla_a f_b - \nabla_b f_a = \partial_a f_b - \partial_b f_a = 0.
\ee
Even if this condition is satisfied though, the force might not be globally a gradient, igniting topological steady currents. Notice that the covariant condition is equivalent to the noncovariant condition (second passage) by virtue of the torsionlessness of the connection. We are tempted to speculate that torsion might therefore be a nonequilibrium drive, but we leave this to future inquiry.

The entropy production can be expressed in terms of the time derivative of the relative entropy, which quantifies the internal entropy production, and an additional environmental term, $\sigma = - \dot{D}_{\sqrt{g}} + \left\langle A^a f_a \right\rangle$, where
\be
\left\langle A^a f_a \right\rangle =  \int p(dx) \, f^a  \left(f_a + \partial_a  \log  \sqrt{g} - \partial_a \log  p \right).
\ee
The second term between curved parenthesis can be shifted from the environmental to the internal term, returning the standard Gibbs-Shannon entropy and the standard environmental term usually considered in the literature \cite{esposito}. The covariant splitting of the entropy production is one of many possibilities that are equivalent up to the choice of reference prior, which is an additional fundamental symmetry of nonequilibrium thermodynamics, deeply mingled with coordinate covariance.

\section{\label{comparison}Comparison of models}

We consider several models of state-dependent diffusion with vanishing exernal forces that can be found in the literature, and compare them with (A) our model: (B) Graham's covariant theory of nonequilibrium statistical mechanics, (C) Diffusion in nonuniform temperature, (D) Diffusion with state-dependent viscosity. In the overdamping limit (B) and (C) lead respectively to Stratonovich and It\=o pure SDE's. For sake of completeness, we also consider (E) the pure SDE with  posticipatory character (Klimontovich convention). Finally we include two more general theories: (F) a noncovariant theory with state-dependent viscosity and temperature, and (G) its covariant counterpart, a theory where the diffusion matrix and the metric need not necessarily coincide. The latter includes both our theory and diffusion with state-dependent viscosity as special cases. In \ref{appendix2} and \ref{appendix3} the overdamping limit for the latter two theories are performed. All others follow as special cases. We resume all these theories' properties in tables \ref{tab:1} and \ref{tab:2} and through the interdependence diagram in \ref{compa}. Notice that (G) preceeds (D) and (F) in the discussion. In this section we will set $ (2 \pi \hbar)^2 /  mc^2 = 1$.

\subsection{(B) Second order Graham theory}

Suppose to replace $F_1$ by $F_2$  in our theory. By taking the overdamping limit, along the same lines exposed in  \ref{appendix2}, one would obtain Graham's first-order theory \cite{graham}. Therefore, we can already conclude that the theory is coordinate covariant but it is not gauge invariant. The derivation leads to the following Stratonovich pure differential equation
\be
\dot{x}^a = \sqrt{2} \, e^a_i \circ \zeta^i =  \delta^{ij} e^b_i \partial_b e^a_j + \sqrt{2}\,e^a_i  \zeta^i . \label{eq:stratoSDE}
\ee
The corresponding manifestly covariant FP equation is
\be
\dot{\rho} = \nabla_{\!a} ( \nabla^a \rho -  \omega^a \rho),
\ee
where we remind that $\omega^a$ is the anholonomicity vector, defined in Eq.(\ref{eq:anholonomicity}). As expected, the theory is not gauge invariant since the anholonomicity vector is not. Hence, knowledge of the covariance matrix $g^{ab}$ is not sufficient for determining the theory. All statistical properties depend on the choice of frames $e_a^i$, including the equilibrium/nonequilibrium character. If for some choice of frame the theory satisfies detailed balance, $\omega_a = - \nabla_a \phi$, since the inhomogeneous term in Eq.(\ref{eq:gaugetrans}) is not generally a gradient, the gauge transformed theory might not satisfy detailed balance. As an example, consider a two-dimensional theory with $e^a_i = \delta^a_i$ and $\omega_a = 0$ in some gauge and coordinates. In two dimensions $R_i^j(x) = R_i^j(\vartheta(x))$ is a rotation matrix by an angle $\vartheta(x)$. To first order in $\vartheta$ the transformed anholonomicity vector $(\tilde{\gamma}_1,\tilde{\gamma}_2 ) = (- \partial_2 \vartheta,\partial_1 \vartheta)$ is certainly not a gradient.

This problem with Graham's theory was also recognized in Ref. \cite{rumpf}.

\subsection{(C) State-dependent temperature} To our knowledge, theories with state-dependent diffusion coefficient and uniform viscosity have been considered in the literature only in one dimension  \cite{matsuo,bringuier,celani,vankampenIBM}, where they have been used as models of systems with varying temperature. Let us generalize to $n$ dimensions:
\be
\ddot{x}_t^a = - \gamma \dot{x}_t^a + \sqrt{2\sigma}\, e^a_i(x_t) \zeta^i_t.
\ee
In this quite general case temperature might not be isotropic, besides being inhomogeneous: It depends on the direction of motion of the Brownian particle. Given the considerations developed in this paper, the theory is gauge invariant but it is not coordinate covariant. The corresponding Kramers generator
\be
\dot{P} = - v^a \frac{ \partial  P}{\partial x^a}  + \frac{ \partial }{\partial v^a} \big( \gamma v^a P \big) + \sigma g^{ab}   \frac{ \partial^2 P}{\partial v^a \partial v^b} \label{eq:tempkram}
\ee
does not admit a locally Maxwellian steady state, as one can appreciate by plugging $\exp [- h_{ab}(x) v^a v^b /2 + k(x)]$ as an \textit{ansatz}. First and third order terms in the velocities can only vanish if $k = 0$ and $\partial_a g_{bc} = 0$. The exact form of the steady state of Eq.(\ref{eq:tempkram}) is not easily computable to our knowledge.

As for Graham's theory, in general the steady state does not satisfy detailed balance. The nontrivial overdamping limit is performed in detail in \ref{appendix3}, leading to the Fokker-Planck equation in It\=o form
\be
\dot{p} = - \partial_a j^a= \partial_a \partial_b \left(g^{ab} p \right) .
\ee
The steady current is $j^\ast_a =  - p^\ast ( g_{ad} \partial_c g^{cd} + \partial_a \log p^\ast )$. For it to vanish, hence for the steady state to be equilibrium, it is necessary that $g_{ad} \partial_c g^{cd}$ be an exact differential. However, this is not generally the case. Detailed balance does hold when temperature is isotropic, by which we mean that there exist coordinates such that
\be
g_{ab}(x) \propto T(x)^{-1} \delta_{ab},
\ee
In this case the metric is said to be conformally flat.

In a parallel publication we argue on physical grounds that our model should be considered as a valid alternative to diffusion in nonuniform temperature \cite{temp}.

\subsection{(G) Metric/diffusion theory} 

One peculiarity of our theory is that the inverse diffusion matrix coincides with the metric: Noise determines the form of deterministic geodesic motion. This provides an elegant theory, but it is not strictly mandatory for a generally covariant theory.

We might introduce a new  nondegenerate metric $h_{ab}$, with Christoffel coefficients $\Omega^a_{bc}$. Let the diffusion matrix $g^{ab}$ be a twice covariant symmetric tensor, possibly different from the inverse metric $h^{ab}$. The equation
\be
\ddot{x}^a + \Omega^a_{bc} \, \dot{x}^b \dot{x}^c = - \gamma^a_b \dot{x}_t^b  + \sqrt{2\sigma }\, e^a_i  \zeta^i,
\label{eq:tensor}
\ee 
with $g^{ab} = \delta_{ij} e^i_a e^j_b$, is generally covariant when $ \gamma^a_b$ is a tensor. We assume the Einstein relation $ \gamma^a_b = \beta \sigma g^{ac} h_{cb}$. The diffusion matrix  and the metric  govern respectively the noisy and the deterministic behavior. The viscosity coefficient   is a tensorial object, once covariant and once contravariant. It is a hybrid of the diffusion matrix and of the metric, interpolating between stochastic and deterministic behavior.  The Kramers equation affords the equilibrium steady state
\be
P^\ast = h \exp -  h_{ab} v^a v^b /2 
\ee
Notice that the metric, and not the diffusion matrix, determines the form of the steady state. Our theory is recovered  by setting $h_{ab} = g_{ab}$. The overdamping limit is performed in \ref{appendix2}, yielding
 \be
\dot{p} =  \partial_a  \left[  h^{ac} g_{cd} h^{db} \left( \partial_b p  - p \,\partial_b \log \sqrt{h}\right)\right].  \label{eq:tensorfokker}
 \ee
 
\subsection{(D) State-dependent viscosity}

Theories with state-dependent diffusion and viscosity are widely studied.   See Ref.\cite{lau} for a review. The $n$-dimensional case, with a rigorous treatment of the overdamping limit, is discussed in Ref.\cite{sanchodurr}, while Ref.\cite{hagesawa} gives a microscopic derivation in the spirit of the Caldeira-Leggett model and Refs.\cite{arnold1,arnold2} discuss the path integral formulation.

The free model is a special case of the above tensor viscosity model when $h_{ab} = \delta_{ab}$ and $\Omega^a_{bc}= 0$. This identification can only hold in one given set of coordinates, hence the theory is not covariant. Although, it is gauge invariant. The steady state is the usual Maxwell-Boltzmann state, displaying no dependence on the spatial variable: In a way, it is blind to the inhomogeneity of the medium.  In the overdamping limit, the theory leads to the SDE
\be
\dot{x}_t^a = \partial_b g^{ab}  + \sqrt{2}\, e^a_i \zeta^i_t . \label{eq:Klimo}
\ee
The Fokker-Planck generator is $\dot{p} = \partial_a \left( g^{ab} \partial_b p \right)$, sometimes named after  Fourier and Fick. The steady state always satisfies detailed balance. In one dimension Eq.(\ref{eq:Klimo}) reduces to a pure SDE in Klimontovich interpretation. 

\subsection{(E) Pure posticipatory SDE} Theories (B) and (C) led respectively to the Stratonovich and to the It\=o differential in the overdamping limit. For sake of completeness, we also consider the pure SDE in Klimontovich form:
\be
\dot{x}^a  =  \sqrt{2}\,e^a_i(x_t + dx_t)  \zeta^i =   2\, \delta^{ij} e^b_i \partial_b e^a_j + \sqrt{2}\,e^a_i(x_t)  \zeta^i.
\ee
The theory is neither covariant nor gauge invariant. Moreover, the drift term arising in the overdamping limit, as for the Stratonovich and the It\=o case, is not generally a gradient, so the theory does not accomodate detailed balance as an essential feature. It is remarkable that, in higher dimensions, none of the pure stochastic differentials generally affords equilibrium steady states. 

\subsection{(F) State-dependent viscosity and temperature.} To complete the collection, we recall that Van Kampen had considered a 1-dimensional generalization of state-dependent temperature and state-dependent viscosity to the case where both are state dependent \cite{vkdiffin}. Jayannavar and Mahato \cite{mahato} claimed to have derived the theory from a microscopic model of state-dependent interactions with a bath, although it appears that the state-dependence of the temperature is assumed along the way, rather than being derived. Properties of the theory are reported in the resuming tables in \ref{compa}.

\section{\label{conclusions}Conclusions}

There are two viable perspectives on the results hereby derived. On the one hand, they might be seen as an exercise on abstract symmetry principles, later to be applied to theories that already enjoy diffeomorphism invariance, such as physics on curved spacetimes and General Relativity.  We point out that relativistic diffusion is a broadly studied subject, and that it has some crucial peculiarities that do not allow direct application of our results. In this respect, although our Langevin equation is similar to Eq.(51) in Ref.\cite{acosta} (but for a term $2 \Gamma^a_{ab} v^b P$), their meaning is very different as to the role of time and the relationship between the diffusion matrix and the metric. Still, covariant diffusion has an interest on its own. Graham introduced his own theory with these words: ``Physical properties are independent of the coordinates used. Hence they must be formulated in terms of covariant quantities''  \cite{graham}. The will to ``covariantize'' (notions of) nonequilibrium thermodynamics was the ultimate drive for the present work as well. Recently Smerlak theorized that, along with propagation, it might be possible to ``tailor'' the diffusion of light in suitably engineered metamaterials \cite{smerlak}. Our proposal and his belong to the same  paradigm, that spatially-dependent diffusion coefficients might act as an effective curvature of space(time). 

On the other hand, our work might be seen as a doorway to state-dependent diffusion even for mesoscopic systems, both at the level of interpretation and for physical modeling. By no means do we claim that the generally covariant Langevin equation is \textit{the} correct equation for state-dependent diffusion, but it should be in all situations where the statistical properties of noise do not dependent on one preferred coordinate system. This demanding hypothesis has to be addressed independently, for example when performing microscopic derivations. A class of systems where curved Brownian motion is relevant is protein transport inside and across cell membranes under the effect of membrane curvature (see the introduction of Ref.\cite{castro2} and references therein). In a parallel publication we also argue that it might be relevant for Brownian motion in non-uniform temperature environments \cite{temp}. The latter conjecture is already testable, as experimental works by Lan\c con et al. \cite{lancon} and by Volpe, Brettschneider et al. \cite{volpe} (after a few issues have been resolved \cite{fog}) display great control over Brownian particles under the effect of spatial dependence of the diffusion coefficient, in this case due to the presence of walls. It might be feasible to design similar experiments where spatial dependence is due to temperature gradients or both to temperature gradients and spatial inhomogeneity, and put several of the models we described above at work. The generally covariant Langevin equation could also find application in Metropolis Langevin Monte Carlo sampling methods, according to the lines of research developed in Ref.\cite {girolami}.

The relationship between gauge invariance and detailed balance was a main theme in this article. It follows from our discussion that gauge invariance is necessary (but not sufficient) to write thermodynamic theories of state-dependent diffusion that enjoy equilibrium steady states, as different choices of gauge will otherwise lead to nonequilibrium driving. This was the case for Graham's theory. Surprisingly, instead, curvature does not affect the equilibrium character of steady states; it might be interesting to consider instead the thermodynamic effect of geometric torsion.

Finally, we point out that most thermodynamic considerations in this work were carried out at the macroscopic level of the Fokker-Planck equation, and regarded only the equilibrium/nonequilibrium character of the steady state and the covariant definition of thermodynamic quantities. A great deal of covariant stochastic thermodynamics, in the spirit of Ref. \cite{imparato}, has yet to be developed.

\paragraph*{Aknowledgments.}

The author is thankful to A. Bazzani for several discussions over the Ph.D. years, and to M. Smerlak for useful comments. The research was partly supported by the National Research Fund Luxembourg in the frame of project FNR/A11/02.

\appendix

\section{Exponential decay to equipartition\label{appendix1}}

We show that average energy has an exponential decay to its steady value, given by the equipartition theorem. 
We consider energy as defined in Eq.(\ref{eq:energy}) --- setting $m \lambda_c^2 = 1$ --- and take its stochastic differential. The key technicality is to use It\=o's Lemma
\be
d H = \frac{\partial  H}{\partial x^a} dx^a + \frac{\partial  H}{\partial v^a} dv^a + \frac{1}{2} \frac{\partial^2  H}{\partial v^a \partial v^b} dv^a  dv^b.\ee
where we only retain contributions of order $dt$. Plugging into the generally covariant Langevin equation we obtain
\be
d H = \frac{\partial  H}{\partial x^a} v^a dt + \frac{\partial^2  H}{\partial v^a \partial v^b}  \delta^{ij} \sigma e^a_i e^b_j dt  
 + \frac{\partial  H}{\partial v^a} \left( - \gamma v^a dt - \Gamma^a_{bc} v^b v^c dt 
+ \sqrt{2\sigma}\, e^a_i dW^i \right).
\ee
Standard manipulations lead  to
\be
\dot{H} =   - 2 \gamma \left(H  -  \frac{n \sigma}{2\gamma} \right)   +  \sqrt{2\sigma}\,  g_{ab} v^a e^b_i \zeta^i .
\ee
Interestingly, the drift term only depends on the energy itself, so that we can average out the noisy term and obtain a closed equation for the average energy $E_t = \langle H_t \rangle$, with solution
\be
E_t  =  e^{-2 \gamma t} \left( E_0 - \frac{n \sigma}{2\gamma} \right) +  \frac{n \sigma}{2\gamma}. 
\ee
It decays exponentially to its steady value $n \sigma / 2\gamma$, which by the equipartition theorem is identified with $n k_B T/2$, yielding the Einstein relation.

\section{Overdamping limit of metric/diffusion theory\label{appendix2}}

We show that the generally covariant stochastic differential equation  Eq.(\ref{eq:gencov}) (Brownian motion on a manifold) and the generally covariant Langevin equation Eq.(\ref{eq:gencovlan}) are equivalent in the overdamping limit $\sigma \to \infty$, at fixed temperature. More precisely, we work with the corresponding diffusion equations, from Eq.(\ref{eq:covlanfp}) to Eq.(\ref{eq:covfpe}). In fact, we prove the result for the more general theory (G) with independent metric and diffusion matrix discussed in Sec.\ref{comparison}, and specialize the result at the very end. The treatment traces out Schuss's \cite[\S8.2]{schuss}. To simplify, solutions to the perturbative chain of equations that will emerge are given, as they are known from previous literature. A logically  close step-by-step derivation can be operated by means of expansions in terms of Hermite polynomials, as is done in Ref.\cite{matsuo}.

Set both $\beta = 1$ and $ (2\pi\hbar)^2 / mc^2 =1$. Define the adimensional energy
\be
\phi(x,v) =  h_{ab}(x) v^a v^b /2.
\ee
Let $\epsilon = \sigma^{-1}$. We rescale time according to $t = \epsilon^{-1} s$ in the covariant Kramers equation Eq.(\ref{eq:covlanfp}), and expand to order $\epsilon$ both the generator and the probability density:
\be
\Big(\epsilon^2 L_2 +  \epsilon L_1 + L_0 \Big) \Big(P_0 + \epsilon P_1+  \epsilon^2 P_2 + \ldots\Big) = 0,
\ee
where
\bes
L_0 P & = & g^{ab} \frac{ \partial }{\partial v^a}  \left( h_{bc} v^c  P + \frac{ \partial P}{ \partial v^b} \right) ,  \\ 
L_1 P & = & - v^a \frac{ \partial  P}{\partial x^a}  + \frac{ \partial }{\partial v^a} \big( \Omega^a_{bc}v^b v^c P \big), \\
L_2 P & = & - \frac{\partial P}{\partial s} .
\ees
We obtain the following expansion:
\bes
0 & = & L_0 P_0 ,  \label{eq:L0} \\
0 & = &  L_0 P_1 + L_1 P_0 ,  \label{eq:L1} \\
0 & = &  L_0 P_2 + L_1 P_1 + L_2 P_0 , \label{eq:L2} \\
& \ldots &   \nonumber
\ees
A solution of the first equation is
\be
P_0(x,v,s) = q_0(x,s) \exp - \phi(x,v),
\ee
where $q_0(x,s)$ is an undetermined function. The spatial density is found by performing the integration over velocities:
\be
p_0(x,s) = \int dv \, P_0(x,v,s) =  \frac{N  q_0(x,s)}{\sqrt{h(x)}}. \label{eq:suppspatial}
\ee
We also define the conditional probability
\be
\pi(v|x) = \frac{P_0(x,v,s)}{p_0(x,s)} = \frac{\sqrt{h(x)}}{N}  \exp -\phi(x,v),
\ee
which is a multivariate normal distribution.
Plugging  $P_0$ into Eq.(\ref{eq:L1}), we obtain
\be
g^{ab} \frac{ \partial }{\partial v^a}  \left( h_{bc} v^c  P_1 + \frac{ \partial P_1}{ \partial v^b} \right) 
 =   \left( v^a \frac{\partial}{\partial x^a} \log \frac{q_0}{h}\right) P_0.
\ee
A solution is given by
\be
P_1 = \left( q_1  - v^a \theta_a \right) P_0,
\ee
where $q_1= q_1(x,s)$ is again an undetermined function and $ \theta_a$ is easily seen to be
\be
\theta_a = g_{ab} h^{bc} \frac{\partial}{\partial x^c} \log \frac{q_0}{h}.
\ee
Substituting into Eq.(\ref{eq:L2}) and integrating with respect to velocities gives
\be
 \frac{\partial p_0}{\partial s}= \frac{\partial}{\partial s}   \int P_0 dv 
= - \int v^a \frac{ \partial P_1 }{\partial x^a}  dv, \label{eq:pass}
\ee
where it is assumed that all terms decay sufficiently fast for high velocities, so that
\be
\int  dv  \, g^{ab} \frac{ \partial }{\partial v^a}  \left( h_{bc} v^c  + \frac{ \partial}{ \partial v^b} \right)  P_2=0, \qquad
\int  dv \frac{ \partial }{\partial v^a} \big( \Omega^a_{bc}v^b v^c P_1 \big) = 0.
\ee
Eq.(\ref{eq:pass}) then yields
\be
 \frac{\partial p_0}{\partial s}  = - \int v^a \frac{ \partial}{\partial x^a}  \left[ \left( q_1  - v^b \frac{\partial}{\partial x^b} \log \frac{q_0}{g}\right) P_0 \right] dv.
 \ee
The first integrand is odd and its contribution vanishes. In view of Eq.(\ref{eq:suppspatial}), we obtain
\be
 \frac{\partial p_0}{\partial s}  =  \frac{ \partial}{\partial x^a}   \left[\left( \int v^a v^b P_0 \, dv \right) g_{bd} h^{dc}  \frac{\partial}{\partial x^c}  \log \frac{p_0}{\sqrt{h}}\right].
 \ee
 We now perform the integration
 \be
 \int v^a v^b P_0(x,v,s) \, dv  =  p_0(x,s) \int v^a v^b \pi(v|x) dv = p_0(x,s) h^{ab}(x), \label{eq:mom2}
 \ee
 where we recognized the covariance matrix of $\pi(\cdot,x)$.
  Finally we obtain
 \be
 \frac{\partial p_0}{\partial s}  =  \frac{ \partial}{\partial x^a}   \left( p_0 h^{ab} g_{bd} h^{dc}  \frac{\partial}{\partial x^c}  \log \frac{p_0}{\sqrt{h}}\right) \ee
In particular, when $h_{ab} =  g_{ab}$ we obtain Eq.(\ref{eq:covfpe}). It follows that in the overdamping limit the generally covariant Langevin equation and brownian motion on a manifold furnish equivalent descriptions, up to first order in $\epsilon$.

\section{Overdamping limit of diffusion with state-dependent temperature and viscosity \label{appendix3}}

We perform the overdamping limit of theory (F), with state-dependent temperature and viscosity. The particular case with state-dependent temperature is obtained by setting  $h^{ab} =  g^{ab}$. The main difference from the above overdamping limit is that the term with Christoffel coefficients disappears, resulting in a minor complication. All goes as before, but for the definition of the operator 
\be
L_1 P = - v^a \frac{ \partial  P}{\partial x^a}.
\ee
The solution to the first equation in the pertubative chain is the same. As above, plugging $P_0$ into Eq.(\ref{eq:L1}), we obtain
\be 
g^{ab} \frac{ \partial }{\partial v^a}  \left(h_{bc}  v^c  P_1 + \frac{ \partial P_1}{ \partial v^b} \right)  
 =   \left(v^a  \frac{\partial}{\partial x^a} \log q_0 - v^a v^b v^c \, \Omega_{abc} \right) P_0. \label{eq:c2}
\ee
We look for solutions in the form
\be
P_1 = \left( q_1 + \frac{1}{3} v^a v^b v^c \Sigma_{abc}  - v^a \theta_a \right) P_0,
\ee
where $q_1= q_1(x,s)$ is again a scalar function, $\theta_a$ is a vector field and $\Sigma_{abc}$ a tensor. After some work we obtain that $\Sigma_{abc} = g_{ad}h^{de} \Omega_{ebc}$ and
\be
\theta_a = g_{ab} h^{bc} \left(\frac{\partial}{\partial x^c} \log q_0 - \frac{4}{3} h^{de}  \Omega_{dea} - \frac{2}{3} h^{de} \Omega_{ade}\right) = g_{ab} h^{bc} \left( \partial_c \log q_0 - \frac{2}{3} \partial_c \log \sqrt{h}  + \frac{2}{3} h_{ad} \partial_e h^{de}\right),  
\ee
make Eq.(\ref{eq:c2}) satisfied, where we used the identities
\be
h^{bc}  \Omega_{bca} = \partial_a \log \sqrt{h} , \qquad
h^{bc} \Omega_{abc} = - h_{ac}  \partial_b h^{bc} - \partial_a \log \sqrt{h} .
\ee
Substituting into Eq.(\ref{eq:L2}) and integrating with respect to $dv$ we obtain
\bea
\frac{\partial p_0}{\partial s} = -\frac{ \partial }{\partial x^a} \int dv \,\left(  -   v^a v^b \theta_b + \frac{1}{3}   v^a v^b v^c v^d g_{bf}h^{fe} \Omega_{ecd} \right) P_0 ,  =  - \frac{ \partial }{\partial x^a} \left(j^a_{1} + j^a_{2} \right) \qquad \qquad
\eea
where again the term with $q_1$ drops for it gives an odd integrand. Given Eq.(\ref{eq:mom2}), the first current contribution reads
\be
 j^a_{1} 
=   p_0 h^{ab}g_{bc}  \left( -  \partial^c \log p_0 - \frac{1}{3} \partial^c \log \sqrt{h}  - \frac{2}{3} \partial_d h^{dc} \right).
\ee
where we raised indices with $h^{ab}$. By the Isserli-Wick theorem
\be
\int dv \, v^a v^b v^c v^d P_0 = p_0 \left( h^{ab} h^{cd} + h^{ac} h^{bd} + h^{ad} h^{bc} \right).
\ee
Letting $G^{ab} = h^{ac} g_{cd} h^{db}$
\be
 j^a_{2}  = \frac{p_0}{3} \left[h^{ab} g_{bc} (-\partial_d h^{dc} - \partial^c \log \sqrt{h} ) +
2  h^{ab} G^{cd} \Omega_{cdb} \right],
\ee
for a total Fokker-Planck current:
\be
j^a = - h^{ab}g_{bc}\left[ \partial^c p_0 + \left(\frac{2}{3} \partial^c \log \sqrt{h} + \partial_d h^{dc} \right) p_0 \right] +  \frac{2}{3}  h^{ab} G^{cd} \Omega_{cdb}  p_0 .
\ee
In particular when $h^{ab} = g^{ab}$ we obtain the It\=o generator $\partial_s p_0  =  - \partial_b \partial_a \left(g^{ab}p_0\right)$, while in the 1-dimensional case we find:
\be
\frac{\partial p_0}{\partial s}  =  - \frac{g}{h^2}\left[ - \partial p_0 + \frac{p_0}{3} \left(2 + \frac{g}{h} \right) \partial \log h \right].
\ee

\section{\label{compa}Comparison of models}

The following diagram summerizes the interdependence of the several Langevin equations that we considered. The symbol $\leadsto$ means ``is replaced by''. 
\be
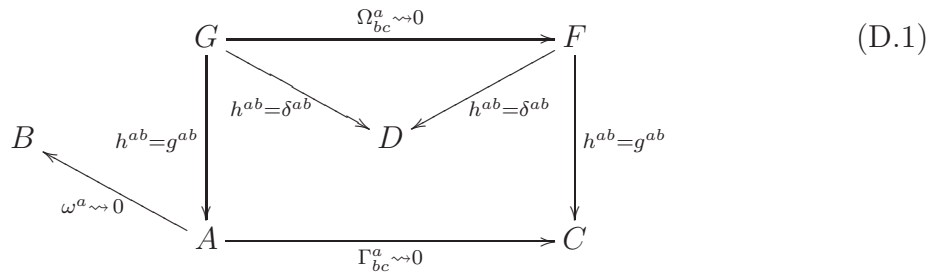
\xymatrix{
& & G  \ar@{->}^{ \Omega^a_{bc} \leadsto 0 }[rrrr]   \ar@{->}_{ h^{ab} = g^{ab} }[dd]  \ar@{->}_{ h^{ab} = \delta^{ab} }[drr] & & & & F  \ar@{->}^{ h^{ab} = g^{ab} }[dd]  \ar@{->}^{ h^{ab} = \delta^{ab} }[dll]  \\
B & & & & D & & \\
& & A \ar@{->}_{ \Gamma^a_{bc} \leadsto 0 }[rrrr]  \ar@{->}^{ \omega^a \leadsto
 \, 0 }[ull]  & & & & C}
\ee 
R\'esum\'e tables can be found in the next page.

\begin{landscape}

    \ctable[
      caption= State-dependent diffusion theories in second-order formalism.,
      label=tab:1,
      captionskip=\smallskipamount]{l|l l l l l}
      {\tnote[$\dagger$]{$\Omega^a_{bc}$ are the Christoffel coefficients of $h^{ab}$}}
{\hline\hline
& \textbf{Model} & \textbf{Refs.} & \textbf{Langevin equation} & \textbf{Einstein relation}  & \textbf{Steady state} \\
\hline \hline
A~ & Generally Covariant & \textit{ivi}, \cite{kleinert,castro2} & $\ddot{x}^a + \Gamma^a_{bc} \dot{x}^b \dot{x}^c = - \gamma \dot{x}^a + \sqrt{2\sigma} e^a_i  \zeta^i$
& $\gamma = \beta \sigma$ & $g \exp - \tfrac{\beta}{2} g_{ab} v^a v^b$ \\
B~ & Covariant (Graham)& \cite{graham,muratore} &  $\ddot{x}^a =  e^i_b \partial_c e^a_i \dot{x}^b \dot{x}^c - \gamma \dot{x}^a + \sqrt{2\sigma} e^a_i  \zeta^i$ &  $\gamma = \beta \sigma$ & ?  \\
C~ & State Dep. Temperature & \cite{sekimotobook,matsuo,bringuier,celani,vankampenIBM}  & $\ddot{x}^a  = - \gamma \dot{x}^a + \sqrt{2\sigma} e^a_i  \zeta^i$ &
$\gamma = \beta \sigma$ & ? \\
D~ & State Dep. Viscosity  & \cite{sanchodurr,sancho,hagesawa,arnold1} & $\ddot{x}^a =  - \gamma^{a}_b \dot{x}^b + \sqrt{2\sigma} e^a_i  \zeta^i$ &
$\gamma^a_b = \beta \sigma g^{ac}\delta_{cb}$ & $\exp - \tfrac{\beta}{2} \delta_{ab} v^a v^b$   \\
E~ & Pure posticipatory  SDE & \cite[Sec.V]{lau} & --- & --- & ---  \\ 
F~ & St. Dep. Visc. and Temp. & \cite{vkdiffin,mahato}   & $\ddot{x}^a = - \gamma^a_b \, \dot{x}^b_t + \sqrt{2\sigma} e^a_i  \zeta^i$   & $\gamma^a_b  = \beta\sigma g^{ac} h_{cb}$ & ?  \\
G~ & Metric/Diffusion & \textit{ivi} & $\ddot{x}^a + \Omega^a_{bc} \dot{x}^b \dot{x}^c = - \gamma^a_b \, \dot{x}^b_t + \sqrt{2\sigma} e^a_i  \zeta^i$  \tmark[$\dagger$] & $\gamma^a_b  = \beta\sigma g^{ac} h_{cb}$ &  $h \exp - \tfrac{\beta}{2} h_{ab} v^a v^b$  
\vspace{0.2cm} 
}    
 
    \ctable[
      caption=  State-dependent diffusion theories in first order formalism. Symmetries and thermodynamic properties.,
      label=tab:2,
      captionskip=\smallskipamount]{l | l l l | l l l }
    {\tnote[a]{Steady state in the 1-dimensional case.}
    \tnote[b]{Detailed Balance, independently of the specific choice of metric, coodinates and gauge. Specific models might obey detailed balance.}
\tnote[$\ddagger$]{$G^{ab} =  h^{ac} g_{cd} h^{db}$.}
}
{\hline\hline
& \textbf{Smoluchovski eq.}&  \textbf{ FP current}   & \textbf{1D S.S.}\tmark[a] & \textbf{Gauge inv.} & \textbf{Covariance} & \textbf{D.B.}\tmark[b]   \\
\hline \hline
A~ & $\dot{x}^a = - g^{bc} \Gamma^a_{bc} + \sqrt{2} e^a_i \zeta^i $  & $ -g^{ab} \left( \partial_b \, p -  p \,  \partial_b \log \sqrt{g}\right) $ &  $\sqrt{g}$ & Yes & Yes & Yes  \\
B~ & $\dot{x}^a = \delta^{ij} e^b_i \partial_b e^a_j +  \sqrt{2} e^a_i \zeta^i$  & $ - g^{ab} \partial_b p -\left(\delta^{ij} e^a_i \partial_b e^b_j\right) p$ & $\sqrt{g}$ &  No & Yes & No  \\
C~ &  $\dot{x}^a = \sqrt{2} e^a_i \zeta^i$  & $ - \partial_b (g^{ab}  p)  $  & $g$ & Yes & No & No  \\
D~ &  $\dot{x}^a = \partial_b g^{ab} +  \sqrt{2} e^a_i \zeta^i$ &  $- g^{ab}  \partial_b \, p $  &  $const.$ & Yes & No & Yes \\
E~ & $\dot{x}^a =  2 \, \delta^{ij} e^b_i \partial_b e^a_j + \sqrt{2} e^a_i \zeta^i$ &  $- g^{ab}  \partial_b \, p + \delta^{ij} \left(e^b_j  \partial_b e^a_i  -e^a_i \partial_b e^b_j\right) p $
& $const.$ & No & No & No \\
F ~ &
\multicolumn{2}{l}{
(omitted) $\qquad\quad - h^{ab}g_{bc}\left[ \partial^c p + \left(\frac{2}{3} \partial^c \log \sqrt{h} + \partial_d h^{dc} \right) p \right] +  \frac{2}{3}  h^{ab} G^{cd} \Omega_{cdb}  p $   \tmark[$\ddagger$]}  
 &
$h^{\tfrac{2}{3}} e^{\int \frac{g \partial h}{3 h^2}}$
& Yes & No & No  \\
G~ &  $\dot{x}^a = - g^{bc} \Gamma^a_{bc} -  \partial_b \log \sqrt{h/g}   + \sqrt{2}\, e^a_i \zeta^i$ & $-g^{ab} \left( \partial_b \, p -  p \,  \partial_b \log \sqrt{h}\right)$  &  $\sqrt{h}$ & Yes & Yes & Yes  
\vspace{0.2cm}
}

\end{landscape}


\section*{Bibliography}

\begin{thebibliography}{20}

\bibitem{sekimoto} Sekimoto K, {\it Langevin equation and thermodynamics}, 1998 Prog. Theor. Phys. Supplement \textbf{130} 17.

\bibitem{seifert} Seifert U, {\it Stochastic thermodynamics: principles and perspectives}, 2008 Eur. Phys. J. B \textbf{64} 423.

\bibitem{esposito} Van den Broeck C and Esposito M, {\it Three faces of the second law. II. Fokker-Planck formulation}, 2010 Phys. Rev. E \textbf{82} 011144.

\bibitem{hanggi}  H\"anggi P, Talkner P and Borkovec M, {\it Reaction-rate theory: 50 years after Kramers},  1990 Rev. Mod. Phys. \textbf{62}  251.

\bibitem{lau} Lau A W C and Lubensky T C, {\it State-dependent diffusion:  Thermodynamic consistency and its path integral formulation}, 2007 Phy. Rev. E \textbf{76} 011123.

\bibitem{klimontovich1} Klimontovich Y L, {\it Nonlinear brownian motion}, 1994   Phys.-Usp.  \textbf{37} 737.

\bibitem{hsu}  Hsu E P, 2002 \textit{Stochastc Analysis on Manifolds} (Providence: AMS). 

\bibitem{stroock} Stroock D W, 2000 \textit{An Introduction to the Analysis of Paths on a Riemannian Manifold} (Providence: AMS).

\bibitem{vankampen} Van Kampen N G, {\it  Brownian motion on a manifold}, 1986 J. Stat. Phys. \textbf{44} 1.

\bibitem{graham} Graham R, {\it Covariant formulation of non-equilibrium statistical thermodynamics}, 1977 Z. Physik B \textbf{26} 397.

\bibitem{muratore} Muratore-Ginanneschi P, {\it Path integration over closed loops and Gutzwiller's trace formula}, 2003 Phys. Rep. \textbf{383} 299.

\bibitem{geometric1}  Obata T, Hara H and Endo K, {\it Differential geometry of nonequilibrium processes}, 1992 Phys. Rev. A \textbf{45} 6997.

\bibitem{crooks} Crooks G E, {\it Measuring thermodynamic length}, 2007 Phys. Rev. Lett. \textbf{99} 100602.

\bibitem{ren} Ren J, H\"anggi P and Li B, {\it Berry-phase-induced heat pumping and its impact on the fluctuation theorem}, 2010 Phys. Rev. Lett. \textbf{104} 170601. 

\bibitem{geometric2} Chernyak  V Y, Klein J R and Sinitsyn N A, {\it Algebraic topology and the quantization of fluctuating currents}, 2011 {\it Preprint} arXiv:1204.2011.

\bibitem{qian} Jiang D-J, Qian M and Qian M-P, 2004 \textit{Mathematical
Theory of Nonequilibrium Steady States} (Berlin: Springer).

\bibitem{polettini} Polettini M, {\it Nonequilibrium thermodynamics as a gauge theory}, 2012 Eur. Phys. Lett. \textbf{97} 30003.

\bibitem{temp} Polettini M, {\it Constrained Brownian motion as thermodiffusion}, 2012 {\it Preprint} arXiv:1211.6580.
 
\bibitem{sanchodurr} Sancho J M, San Miguel M and D\"urr D, {\it Adiabatic elimination for systems of Brownian particles with nonconstant damping coefficients}, 1982 J. Stat. Phys. \textbf{28} 291.

\bibitem{sancho} Sancho J M, {\it Brownian colloidal particles: Ito, Stratonovich or a different stochastic interpretation}, 2011 Phys. Rev. E  \textbf{84} 062102. 

\bibitem{lancon} Lan\c con P, Batrouni G, Lobry L and Ostrowsky N, {\it Drift witout flux: Brownian walker with a space dependent diffusion coefficient}, 2001 Europhys. Lett. \textbf{54} 28. 

\bibitem{volpe}  Volpe G, Helden L, Brettschneider T, Wehr J and Bechinger C, {\it  Influence of noise on force measurements}, 2010 Phys. Rev. Lett. \textbf{104} 170602; Brettschneider T, Volpe G, Helden L, Wehr J and  Bechinger C, {\it Force measurement in the presence of Brownian noise: Equilibrium-distribution method versus drift method}, 2011 Phys. Rev. E \textbf{83} 041113.
 
\bibitem{fog} Mannella R and McClintock P V E ,  {\it Comment on ``Influence of Noise on Force Measurements''}, 2011 Phys. Rev. Lett. \textbf{107} 078901; Volpe G et al., {\it Reply\ldots}, 2011 Phys. Rev. Lett. \textbf{107} 078902.

\bibitem{sekimotobook}  Sekimoto K, 2010 \textit{Stochastic energetics. Lect. Notes Phys. Vol. 799} (Berlin: Springer).

\bibitem{matsuo} Matsuo M and Sasa S, {\it Stochastic energetics of nonuniform temperature systems}, 2000 Phys. A \textbf{276} 188.

\bibitem{bringuier} Bringuier E and Bourdon A, {\it Colloid transport in non-uniform temperature}, 2003 Phys. Rev. E \textbf{67} 011404. 

\bibitem{celani} Celani A, Bo S, Eichhorn R and Aurell E, {\it Anomalous thermodynamics at the micro-scale}, 2012 {\it Preprint}  arXiv:1206.1742.

\bibitem{vankampenIBM} Van Kampen N G, {\it Relative stability in nonuniform temperature}, 1988 IBM J. Res. Dev. \textbf{32} 107.

\bibitem{landauer} Landauer R, {\it Motion out of noisy states}, 1988 J. Stat. Phys. \textbf{52} 233.

\bibitem{vankampen2} Van Kampen N G, {\it Ito vs. Stratonovich}, 1981 J. Stat. Phys. \textbf{24} 175.

\bibitem{klimontovich2} Klimontovich Y L, {\it Ito, Stratonovich and kinetic forms of stochastic equations}, 1990 Physica A: Stat. Theor. Phys. \textbf{163} 515Ð532.
  
\bibitem{sokolov} Sokolov I M, {\it  Ito, Stratonovich, H\"anggi and all the rest: The thermodynamics of interpretation}, 2010 J. Chem. Phys. \textbf{375} 359.

\bibitem{yuan} Yuan R and Ao P, {\it Beyond Ito vs. Stratonovich}, 2012  J. Stat. Mech.  P07010.

\bibitem{mannella} Mannella R and McClintock P V E, {\it Ito vs. Stratonovich: Thirty years later}, 2012 Fluct. Noise Lett. \textbf{11} 1240010.

\bibitem{muratore2} Muratore-Ginanneschi P, {\it On the use of stochastic differential geometry for non-equilibrium thermodynamics modeling and control}, 2012 {\it Preprint} arXiv:1210.1133.

\bibitem{kleinert} Kleinert H and Shabanov S V, {\it Theory of Brownian motion of a massive particle in spaces
with curvature and torsion}, 1998 J. Phys. A: Math. Gen. \textbf{31}, 7005. 

\bibitem{castro2} Casta\~neda-Priego R, Castro-Villarreal P, Estrada-Jim\'enez S and M\'endez-Alcaraz J M, {\it Brownian motion of free particles on curved surfaces}, 2012 {\it Preprint} arXiv:1211.5799.

\bibitem{batrouni} Batrouni G G, Kawai H and Rossi P, {\it Coordinate-independent formulation of the Langevin equation}, 1986 J. Math. Phys. \textbf{27} 1646.
 
\bibitem{villarreal} Castro Villarreal P, {\it Brownian motion meets Riemann curvature}, 2010 J. Stat. Mech. P08006.

\bibitem{tupperyang}  Tubber P F and Yang Y, {\it  A paradox of state dependent diffusion and how to resolve it}, 2012 {\it Preprint} arXiv:1204.1590.

\bibitem{objsubj}  Berger J, {\it The Case for Objective bayesian Analysis}, 2006 Bayesian Analysis \textbf{1}, 385; Goldstein M, {\it  Subjective bayesian Analysis: Principles and Practice}, 2006 \textit{ibid.}, Bayesian Analysis \textbf{1} 403.

\bibitem{rumpf} Rumpf H, {\it Stochastic quantization of Einstein gravity}, 1986 Phys. Rev. D \textbf{33} 942.
  
\bibitem{hagesawa} Hasegawa H, {\it  Classical open systems with nonlinear nonlocal dissipation and state-dependent diffusion: Dynamical responses and the Jarzynski equality} 2011 Phys. Rev. E \textbf{84} 051124.

\bibitem{arnold1} Arnold P, {\it Langevin equations with multiplicative noise: resolution of time discretization ambiguities for equilibrium systems}, 2000 Phys. Rev. E \textbf{61} 6091; 

\bibitem{arnold2} Arnold P, {\it Symmetric path integrals for stochastic equations with multiplicative noise}, 2000 Phys. Rev. E \textbf{61}  6099.

\bibitem{vkdiffin} Van Kampen N G,  {\it Diffusion in inhomogeneous media}, 1987 Z. Phys. B \textbf{68} 135.

\bibitem{mahato} Jayannavar A M  and Mahato M C, {\it Macroscopic equation of motion in inhomogeneous media: A microscopic treatment} 1995 PRAMANA  Journal of Physics \textbf{45} 369.


\bibitem{acosta} Chac\'on-Acosta G and Kremer G, {\it Fokker-Planck-type equations for a simple gas and for a semirelativistic Brownian motion from a relativistic kinetic theory}, 2007 Phys. Rev. E \textbf{76} 021201.

\bibitem{smerlak} Smerlak M, {\it Tailoring diffusion in analogue spacetimes} 2012 Phys. Rev. E \textbf{85} 041134. 
 
\bibitem{girolami} Girolami M and Calderhead B, {\it  Riemann manifold Langevin and Hamiltonian Monte Carlo methods}, 2011 J. R. Statistic. Soc. B \textbf{73} 123. 

\bibitem{imparato}  Imparato A and Peliti L, {\it Fluctuation relations for a driven Brownian paticle}, 2006 Phys. Rev. E \textbf{74}, 026106.

\bibitem{schuss} Schuss Z, 2010 \textit{Theory and Applications of Stochastic Processes} (New York: Springer). 



\end{thebibliography}
\end{document}